\documentclass[preprint,12pt, a4paper]{elsarticle}

\usepackage{amssymb}

\usepackage{float}
\restylefloat{table}

\journal{SoftwareX}

\begin{document}

\begin{frontmatter}



\title{SEDBYS: A python-based SED Builder for Young Stars} 


\author{C. L. Davies}

\address{Astrophysics Group, Department of Physics \& Astronomy, University of Exeter, Stocker Road, Exeter, EX4 4QL, UK}

\begin{abstract}
Spectral energy distributions (SEDs) are useful primary and complementary tools in the analysis of observations of young stars. However, the process of collating, inspecting, and flux-converting archival photometry and spectroscopy to build spectral energy distributions for young stars is time-consuming. Here, I present SEDBYS (Spectral Energy Distribution Builder for Young Stars), a python-based repository of command-line tools built to (i) query online photometric and spectroscopic catalogs and a distributed database of archival photometry, (ii) use a look-up table of zero points to flux-convert the acquired data, (iii) enable interactive visual inspection of the SED and (iv) handle book-keeping to collate references in bibTeX format. The code is distributed via git and is equipped with additional tools to enable users to add existing or forthcoming catalogs to the list of sources queried, ensuring the longevity of SEDBYS as a tool for the star formation community.
\end{abstract}


\end{frontmatter}

\section*{Required Metadata}
\label{}

\section*{Current code version}
\label{}


\begin{table}[H]
\begin{tabular}{|l|p{6.5cm}|p{6.5cm}|}
\hline
\textbf{Nr.} & \textbf{Code metadata description} & \textbf{Please fill in this column} \\
\hline
C1 & Current code version & v2.0 \\
\hline
C2 & Permanent link to code/repository used for this code version & $https://gitlab.com/clairedavies/sedbys/-/releases/v2.0$ \\
\hline
C3 & Code Ocean compute capsule & \\
\hline
C4 & Legal Code License   & BSD 3-Clause "New" or "Revised" license (BSD-3-Clause) \\
\hline
C5 & Code versioning system used & git \\
\hline
C6 & Software code languages, tools, and services used & python \\
\hline
C7 & Compilation requirements, operating environments \& dependencies & Astropy and other standard python libraries; Python $\geq3.5$; Linux, OSX, Windows \\
\hline
C8 & If available Link to developer documentation/manual & For example: $https://gitlab.com/clairedavies/sedbys/-/blob/master/README.md$ \\
\hline
C9 & Support email for questions & c.davies3@exeter.ac.uk \\
\hline
\end{tabular}
\caption{Code metadata (mandatory)}
\label{} 
\end{table}



\section{Motivation and significance}
\label{sec:intro}

%
%
%

The shape of the spectral energy distribution (SED) of a star changes during its formation. Resembling a cool black body in the earliest stages of gravitational collapse, the initially spherical circumstellar envelope is flattened into a disk via the conservation of angular momentum. During this time, the stellar and circumstellar portions of the SED become more distinguished \citep{Lada84, Beall87, Wilking89}. In the latter stages of this evolutionary sequence, the SED resembles stellar blackbody emission plus excess emission from UV to radio wavelengths, arising as a result of the accretion of high energy circumstellar material onto the star, and through the re-processing and scattering of starlight by circumstellar dust and gas \citep{Bertout89, Fischer11}. Analysing the SED of a young stellar object (YSO) can thus reveal its evolutionary status \citep[e.g.][]{Furlan16, Robitaille17}. 

Moreover, studying different portions of the SED of a YSO can shed further light on the nature of the circumstellar material. For instance, the shape of the SED around the $10\,\mu$m silicate feature probes the level of dust grain growth, settling \citep[e.g.][]{DAlessio06}, and processing into crystallised structures \citep[e.g.][]{Sargent09}; a relative dearth of near-infrared emission compared to far-infrared or submillimetre emission can reveal the presence of (potentially planet-carved) annular dust gaps and/or inner disk cavities \citep[e.g.][]{Espaillat12}; and the spectral index across the millimetre to radio portions of the SED can be used to distinguish thermal continuum from free-free emission \citep{Wright75, Eisner08}, enabling the circumstellar dust mass to be estimated \citep[e.g.][]{Eisner18}. Information about the central star may also be inferred from the SED: stellar effective temperatures, luminosities, radii, and surface gravities are commonly estimated by comparing the SED to stellar evolutionary models and/or model atmospheres \citep[e.g.][]{Leggett01, Filippazzo15, Pascual16, Lodieu18, Zhang18}.

SEDs are also vital complementary tools for the analysis of high angular resolution observations of YSOs. Near- and mid-infrared interferometric studies of YSOs often collate and use SEDs to provide an independent assessment of the stellar flux contribution to the wavelength range probed by the high angular resolution observations. This stellar flux contribution is otherwise degenerate with the characteristic size of the circumstellar emitting region \citep{Lazareff2017lg}. High contrast imaging \citep[e.g.][]{Rich15} and interferometric studies \citep[e.g.][]{Davies18} with limited wavelength coverage often also compare their models to SEDs to ensure their disk models are consistent with observations across all annuli.

YSO SEDs are typically compiled from archival data, sometimes supplemented with new photometric observations. Existing tools such as the VizieR Photometry Viewer\footnote{$http://vizier.unistra.fr/vizier/sed/$} or the Virtual Observatory SED Analyser (VOSA \citep{Bayo08}) are extremely useful in helping to collate data from surveys. However, they each suffer from at least one of the following caveats: 
\begin{itemize}
    \item To be collated, photometry must exist in online catalogs. As such, potentially useful archival photometric data which was, for instance, published prior to the development of online catalogs or reported in the main body of a paper rather than being tabulated is completely missed. This particularly effects - but is by no means limited to - long wavelength (millimetre and radio) data. In addition, the converting of data tables in published journals to VizieR cataloged tables is by no means complete, further limiting the data that is retrieved.
    \item Data collation is based on target coordinates rather than individual object identifiers. As such, data for a number of individual stars in a multiple system (or otherwise crowded field) may be collated and combined with blended photometry obtained on telescopes with a wider field of view. 
    \item Details of the observation date and effective telescope resolution or interferometric beam size are not collated, both of which are useful to assess when data for a specific target need to be cleaned due to e.g. a vertical spread in magnitude or flux at any given wavelength.
    \item While the source of the data is retained to aid book-keeping and referencing, some catalogs also contain data which has been sourced from pre-existing catalogs or studies. Other catalogs are simply collations of many different individual sources of photometry. In these cases, the original references are not necessarily retained and extra time and effort must be spent on ensuring the work of the original authors is acknowledged. 
\end{itemize}
Thus, even with these existing tools, collating and inspecting data to build an SED for an individual YSO can often be a time-consuming process. 

Herein, I present the Spectral Energy Distribution Builder for Young Stars (SEDBYS), a python-based git repository\footnote{$https://gitlab.com/clairedavies/sedbys$} of command-line tools designed for the rapid collation of archival photometric and spectroscopic data and the building and inspection of SEDs. The aims of SEDBYS were to address the caveats outlined above, thereby greatly reducing the time taken to compile SEDs from days, if not weeks to mere minutes. In particular:
\begin{itemize}
    \item Create a searchable (and expandable) repository of published photometry and flux measurements which do not already exist in online catalogs, thereby increasing the legacy value of existing photometric survey data. 
    \item Retain the observation date alongside the photometry and flux measurement data. As YSOs are intrinsically variable at optical and infrared wavelengths (see \citep{Cody18} and references therein), this allows the user to prioritise (semi-)contemporaneous photometry and/or a single bright epoch when inspecting their SED. 
    \item Retain a record of the telescope resolution or extent of the interferometric beam alongside the photometry and flux measurements. This information is useful when a target is located in a crowded field or is a component of (or conversely comprises multiple members of) a binary or young multiple stellar system. For longer wavelength observations, the resolution of the observations is also important to consider whether any surrounding nebulosity may contaminate the flux measurement or has instead been resolved-out. 
    \item Provide an interactive plotting tool which can be used in conjunction with the observation date and resolution information to inspect the collated data. This must remain synchronous with the collated data so that a record can be kept of any rejected data, aiding reproducibility of results.
    \item Equip the user with a tool to create LaTeX format, fully referenced data tables and corresponding bibTeX files from the collated data. 
\end{itemize}


%

\section{Database queries}
SEDBYS command line operator {\tt queryDB.py} is designed to search for photometric and spectroscopic data corresponding to a user-specified target young stellar object. In particular, the user specifies the target using its name. The search conducted by {\tt queryDB.py} comprises three stages:
\begin{enumerate}
    \item interface with VizieR to query a select list of online survey data and catalogs for extant photometry;
    \item query the SEDBYS photometry database for extant photometry;
    \item (optional) interface with the Cornell Atlas of Spitzer/IRS Sources (CASSIS \citep{cassis} and Gregory C. Sloan's Atlas of full-scan AOT1 spectra \citep{sws} to search for extant flux-calibrated, low resolution infrared spectra.
\end{enumerate}
The details of these steps are outlined in turn in the following subsections. 

\subsection{Integration with VizieR and SIMBAD}\label{sec:vizcat}
\begin{table}
\begin{tabular}{l|l}
\hline
\textbf{Instrument or Survey} & \textbf{Catalog Reference(s)}\\
\hline
XMM-OM    & \citep{xmm} \\
GALEX     & \citep{galex} \\
Gaia      & \citep{gaia} \\
Tycho-2   & \citep{tycho2} \\
SDSS      & \citep{sdss09, sdss12, sdss15} \\
APASS & \citep{aavso} \\
2MASS     & \citep{2mass} \\
MSX   & \citep{msx} \\
AKARI     & \citep{akariirs, akarifis} \\
SPITZER   & \citep{evans03, caulet08, luhman08, gutermuth09} \\
WISE      & \citep{wise} \\
IRAS      & \citep{iras} \\
Herschel  & \citep{herschel} \\
CSO   & \citep{submm} \\ 
SCUBA     & \citep{scuba, submm} \\
LABOCA    & \citep{laboca} \\
ALMA      & \citep{alma1, alma2, alma3} \\
VLA       & \citep{vla} \\
\hline
\end{tabular}
\caption{Catalogs queried using {\tt astroquery.vizier}}
\label{tab:vizcats}
\end{table}

To interface with VizieR and retrieve extant online photometry across ultra violet to radio wavelengths, SEDBYS employs {\tt astroquery} \citep{Ginsburg19}, an affiliated package of Astropy\footnote{http://www.astropy.org} \citep{astropy:2013, astropy:2018}. The search is restricted to the surveys and catalogs listed in Table~\ref{tab:vizcats} as their format makes it simple to automate the retrieval of measurements of magnitudes or fluxes, together with their measurement uncertainties, in a consistent manner. SEDBYS also provides the user with command line operator {\tt addVizCat.py} which enables this list of surveys and catalogs to be expanded by all users (Section~\ref{sec:addvizcat}).

\begin{figure}
\begin{center}
\includegraphics[width=\columnwidth]{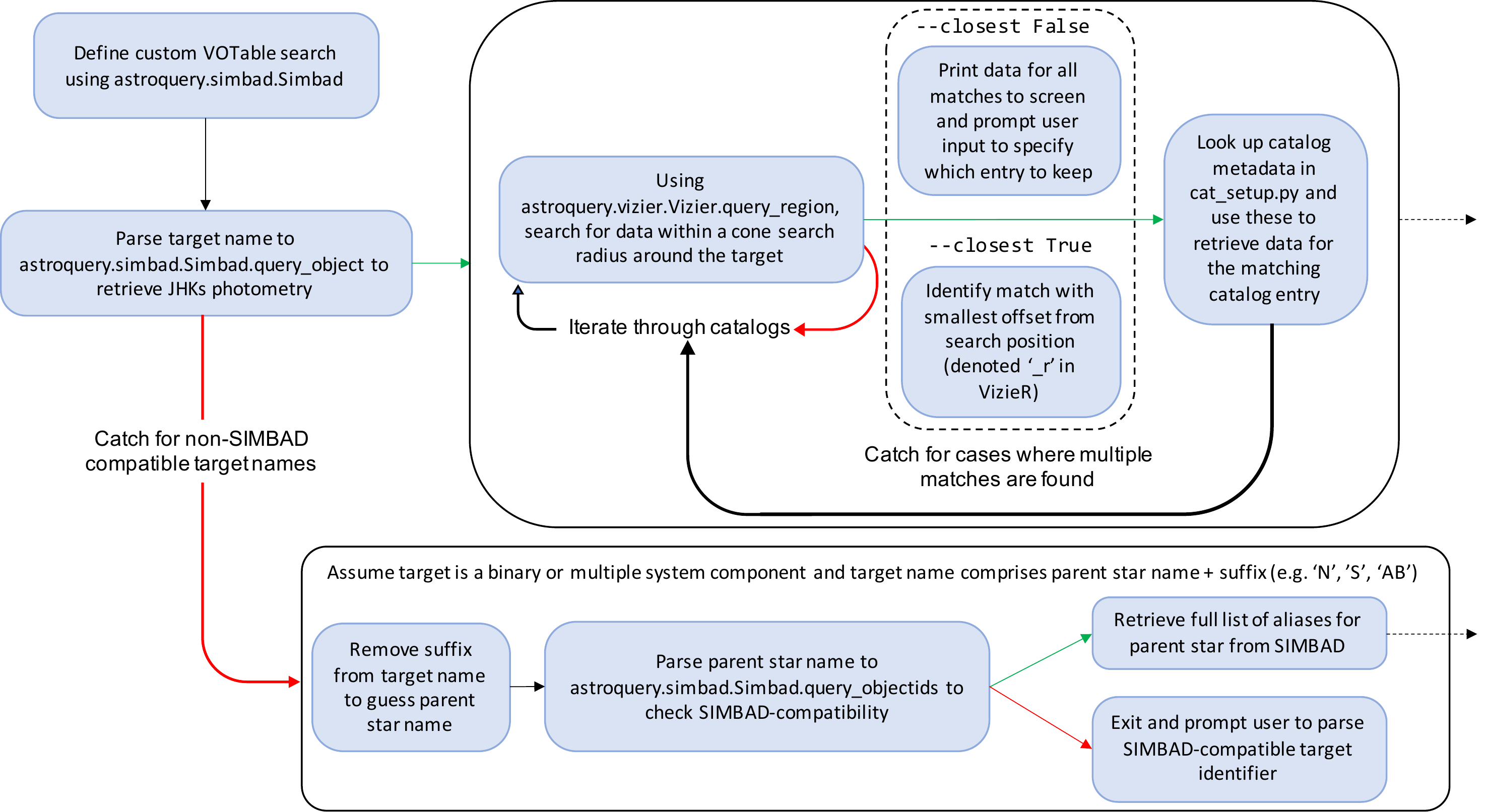}
\caption{A schematic of the first stage of the {\tt queryDB.py} procedure (the interface with VizieR to query the online survey data and catalogs listed in Table~\ref{tab:vizcats}). Green and solid black arrows are used to denote successful paths through the procedure while red arrows denote the path followed when a process is unsuccessful or no matches are found. Dashed black arrows indicate where the procedure continues on to the second stage of {\tt queryDB.py} (Section~\ref{sec:localDB}).}\label{fig:vizcatquery}
\end{center}
\end{figure}

A schematic representation of the procedure followed by {\tt queryDB.py} during step 1 is shown in Fig.~\ref{fig:vizcatquery}. The {\tt astroquery} function {\tt query\_region} uses SIMBAD to convert the target name to sky coordinates and conducts a search for matches within a specified cone radius of these coordinates. A default cone search radius of $10\,$arcseconds is used unless otherwise specified by the user (using optional argument {\tt --rad}).

In some cases, multiple cross-matches may be found within the defined search region. By default, {\tt queryDB.py} operates in interactive mode (using optional argument {\tt --closest False}). In these instances, all potential matches are printed to screen alongside the positional offset\footnote{The positional offset (in arcseconds) between each potential match in the given catalog and the input target sky coordinates is denoted as $\_r$ in VizieR. This symbol is retained in SEDBYS.}, $\_r$, and a message prompting the user to specify which entry corresponds to their target by entering the corresponding $\_r$ value. Alternatively, the user can set optional argument {\tt --closest True} which automatically retrieves the data for the catalog entry with the smallest $\_r$. This option is particularly useful when using {\tt queryDB.py} in batch mode for a large number of YSOs.

The data and catalog metadata\footnote{The catalog metadata is flexible to some of the different ways measurement uncertainties are presented in VizieR. For instance, instead of providing individual measurement uncertainties, \citep{scuba} provide a percentage flux uncertainty. The {\tt queryDB.py} script contains a catch to look for instances of floats rather than strings in the measurement uncertainty catalog metadata entry. Floats are treated as percentage errors and {\tt queryDB.py} combines these with the flux measurement to estimate the uncertainty.} retrieved in this stage are collated into a numpy array before the process progresses to stage 2 (Section~\ref{sec:localDB}).

\subsection{The SEDBYS photometric database}\label{sec:localDB}
\begin{figure}
\begin{center}
\includegraphics[width=0.7\columnwidth]{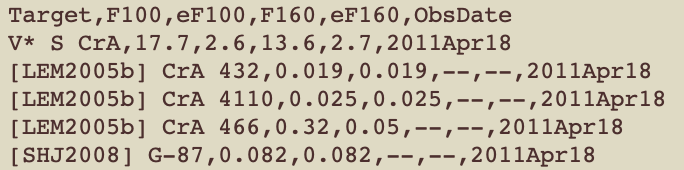}
\caption{Example format of the SEDBYS photometric database files. The first column corresponds to the target name (as it appears in SIMBAD - see text for details) and the final column corresponds to the observation date (in YYYYMmmDD or YYYYMmm format). In between are the flux (or magnitude) measurements and their errors. The catalog metadata in {\tt cat\_setup.py} is used to identify which columns contain measurements, which contain measurement uncertainties, and what the units of these measurements are. Two dashed lines are used to indicate that no measurements are available for a given object in a given waveband or at a given wavelength.}
\label{fig:exampleCSV}
\end{center}
\end{figure}

I have compiled the SEDBYS photometry database from extant publications which are not cataloged in VizieR. All data is collated in a consistent format (Fig.~\ref{fig:exampleCSV}) and catalog metadata is stored in python dictionary format in {\tt cat\_setup.py}. The data are saved alongside the target name. 

The reliance on the target name being SIMBAD-compatible in stage 1 of {\tt queryDB.py} (Section~\ref{sec:vizcat}) is followed through stage 2 as well, albeit with one adaptation. This adaptation accounts for individual components of binary or young multiple stellar systems which often lack their own SIMBAD entry but for which individual component photometry exists in published literature. The identifiers of individual components of binary or multiple systems present in the SEDBYS photometric database (e.g. V* XZ Tau B) comprise a SIMBAD-compatible name for the parent star (e.g. V*~XZ~Tau in this instance) plus a suffix denoting the hierarchy of the target in the multiple system (e.g. `B' in this instance). 

If a user parses a target name to {\tt queryDB.py} which is not SIMBAD-compatible, stage 1 of {\tt queryDB.py} (Section~\ref{sec:vizcat}) is skipped as it would otherwise fail to return any data. Instead, the procedure follows the first red arrow in Fig.~\ref{fig:vizcatquery}. The target name is split into ``parent'' and ``suffix'' components and the ``parent'' component is checked to assess its compatibility with SIMBAD before being used to retrieve all aliases for the parent star. Each alias is then appended with the original suffix to create a list of target aliases to be used in the cross matching process. 

If, instead, the target name parsed to {\tt queryDB.py} is SIMBAD compatible, all aliases are recovered for the target in stage 1. Then, the different files in the SEDBYS database are iterated through to search for target name matches. A partial match may be found on occasion. For instance, if a search is conducted for photometry on a ``parent'' star, the cross-match would locate a partial match when individual component photometry was present in a particular file. In such instances, the data is not retrieved. Instead, a warning message is printed to screen alerting the user that individual component or blended photometry is available in the SEDBYS database and the partial match to the parsed target name is printed to screen. This is intended to alert the user that they may wish to run a separate search on these components.

\begin{figure}
\begin{center}
\includegraphics[width=\columnwidth]{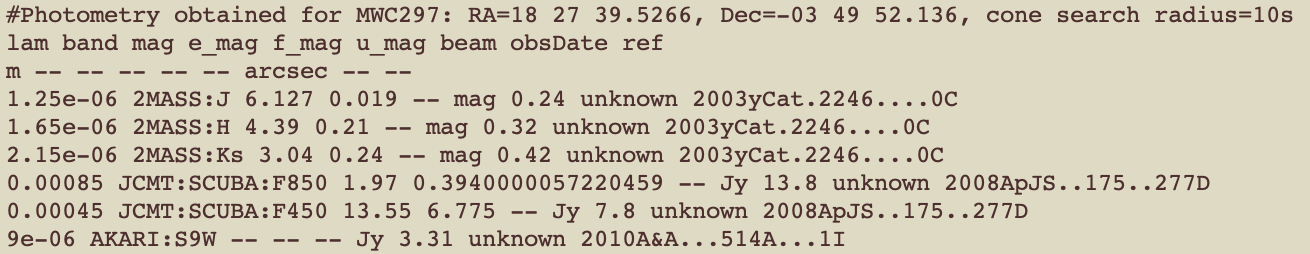}
\caption{First few lines of file {\tt MWC297\_phot.dat} to show an example of the format of the file produced when running {\tt queryDB.py}. The target name and cone radius used in the search, together with the sky coordinates used by {\tt astroquery} are provided in the first line of the file. Then, running left to right, the columns contain the wavelength of the observation, the name of the waveband, the magnitude (or flux), its measurement uncertainty, any flags on this measurement, the units of this measurement, the telescope resolution or effective beam of the observation, the observation date, and the bibliographic reference.}
\label{fig:photdat_example}
\end{center}
\end{figure}

When exact matches are identified, {\tt cat\_setup.py} is used to retrieve the data and catalog metadata which are collated in a numpy array. Only after the stage 2 iteration has reached its conclusion are the data stored in the numpy array written to file. An example of the format of this file is shown in Fig.~\ref{fig:photdat_example}.

\subsubsection{Notes on individual sources of data}
Some of the sources of data which I have compiled into the SEDBYS photometric database used object identifiers which are not SIMBAD-compatible. In these cases, I used the coordinates provided for targets in the source paper to link the data to a SIMBAD-compatible identifier. This also served as a cross-check of previous matching between sky coordinates and object identifiers. It is noted that table 15 in \citep{Ribas2017} presents a naming error for three YSOs in Chamaeleon (namely Hn 11,WY Cha, and T45a; \'{A}. Ribas, private communication; \citep{Ribas20}), which were associated an incorrect name. In SEDBYS, the Herschel fluxes provided in table 15 of \citep{Ribas2017} have been re-matched to their respective target identifiers. Similarly, in \citep{Tripathi17}, the SMA data for target RX~J1633.9-2442 are incorrectly attributed to a target with name RX~J1633.9-2422, which has no match in the ROSAT catalog. This has been corrected in the SEDBYS database.

\subsection{Infrared spectra retrieval}
If the user parses optional argument {\tt --getSpect True}, stage 3 of {\tt queryDB.py} is undertaken. The Cornell Atlas of Spitzer/IRS Sources (CASSIS \citep{cassis}) and Gregory C. Sloan's Atlas of full-scan AOT1 spectra \citep{sws} are queried for low-resolution Spitzer spectra and Infrared Space Observatory Short Wavelength Spectrometer (ISO/SWS) spectra, respectively. The sky coordinates of the target, retrieved from SIMBAD using {\tt astroquery.simbad}, are used when querying CASSIS. In particular, the coordinates retrieved from SIMBAD are converted to decimal degrees using the {\tt coordinates} package of Astropy and concatenated, alongside the cone search radius, into a string matching the format of the CASSIS \emph{Query form}\footnote{$https://cassis.sirtf.com/atlas/query.shtml$} output. The {\tt urlopen} and {\tt urlretrieve} functions of the {\tt urllib.request} python package are then used to (i) scrape the page source to identify the name of any files found by CASSIS which contain low-resolution Spitzer spectra and (ii) download a copy of these files to the user's local machine.

To retrieve the ISO SWS spectra, the {\tt urlopen} function is used to scrape the full page source\footnote{$https://users.physics.unc.edu/~gcsloan/library/swsatlas/aot1.html$}. The sky coordinates of the user-specified search target are compared to those of each target in the atlas until a match within the user-defined cone search radius is found. The {\tt urlretrieve} function is then used to download the corresponding file to the user's local machine. In case an object has been observed multiple times, this search for matches continues until the end of the page source is reached.

To aid book-keeping and referencing, the suggested acknowledgements for CASSIS and Gregory C. Sloan's SWS Atlas are hard-coded into SEDBYS. If data for a target is retrieved from these sources, a bibTeX file is created for the target which includes these acknowledgements.

\section{Data Inspection}\label{sec:inspect}
SEDBYS script {\tt inspectSED.py} allows the user to visually inspect the collated spectroscopic and photometric data. This step is interactive and provides the user with a method to flag any unwanted data while retaining a record of the original search results to aid reproducibility of results. This is particularly useful for instances where, for example, the source may be confused, the data are saturated, or where upper limits are superseded by more sensitive observations. 

The photometric data are converted to flux densities using zero points retrieved from the literature. With the exception of GALEX FUV and NUV and SDSS ugriz data, all measurements are in the Vega system. Python event handling is utilised to provide an interactive plot of the compiled SED. When the data are displayed, the user is prompted to click on any photometric data points they wish to flag for removal. The remaining data (or all the data if none were flagged) are then saved to a separate file containing the flux-converted values. The format of this file is also readable by {\tt inspectSED.py}, enabling the user to repeat this step if necessary. 

The SEDBYS script {\tt toLaTex.py} can then be used to reformat this file into a LaTeX table (see Table~\ref{tab:mwc297_phot} for an example) with corresponding bibTeX file for easy referencing.

\section{Future expansion}\label{sec:addvizcat}
The aim of this first release of SEDBYS has not been to ensure that the local database is complete. Instead, the idea that SEDBYS would be distributed via a git repository has been fed into its development and helps to ensure its growth and longevity. SEDBYS comes equipped with scripts ({\tt addLocal.py} and {\tt addVizCat.py}) which can be used to expand the SEDBYS photometric database and the list of VizieR catalogs to be queried, respectively. Examples of their usage are presented in Section~\ref{sec:eg_additions}. Any updates can then be pushed by the user to the git repository to grow the resource.

\section{Illustrative examples}
\label{sec:example}

\subsection{Database query: young stellar object, MWC 297}
\begin{figure}
\begin{center}
\includegraphics[width=0.8\columnwidth]{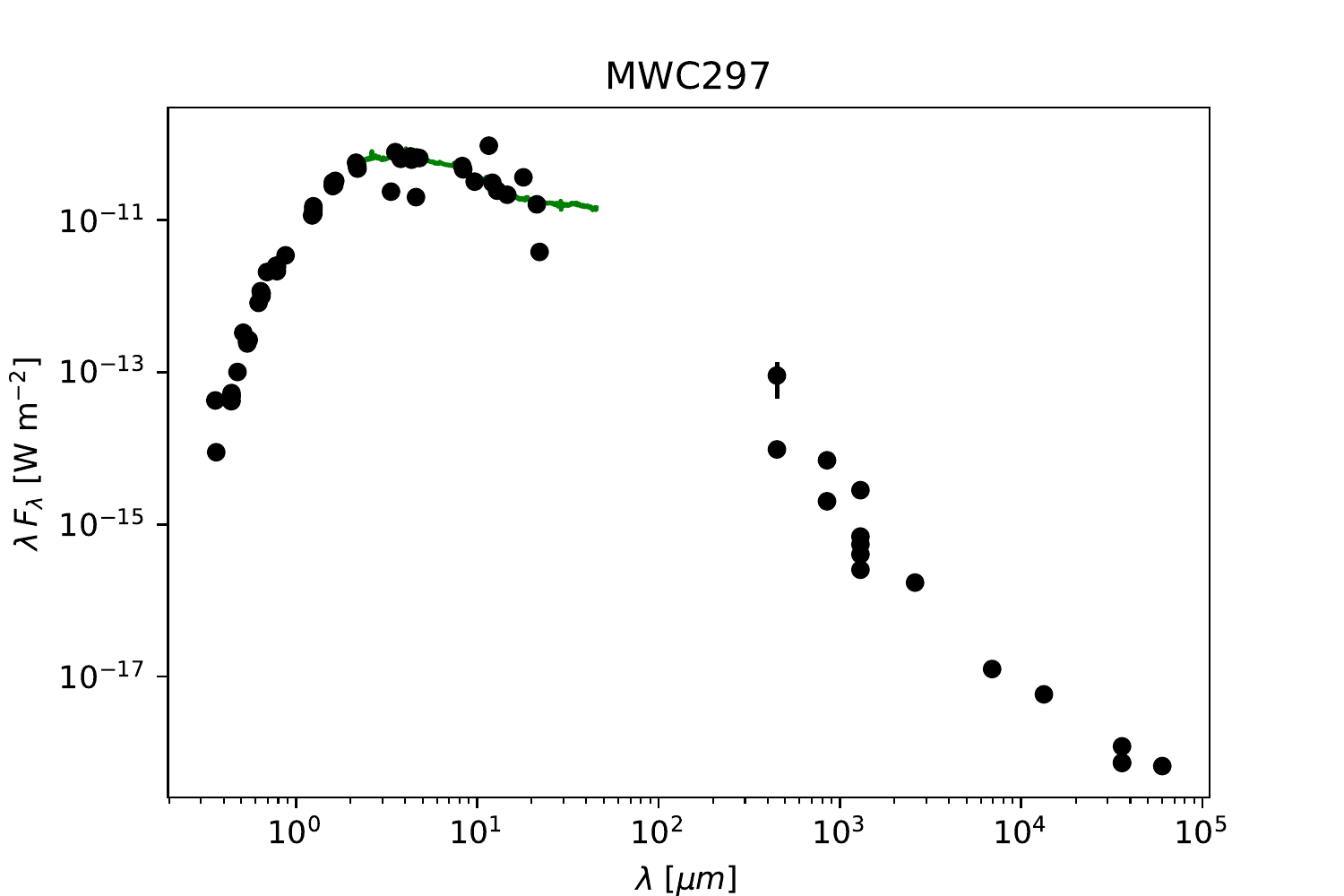}
\caption{SED of MWC~297, built using SEDBYS using the default cone search radius of $10\,$arc seconds with optional arguments {\tt --getSpect True} and {\tt --closest True}. The ISO/SWS spectrum is shown in green while the photometric data are overlaid as black filled circles. In most cases, photometric measurement errors are smaller than the data points.}\label{fig:mwc297}
\end{center}
\end{figure}

Fig~\ref{fig:mwc297} shows the SED compiled using SEDBYS for the YSO MWC~297. This was generated using the following sequence of command-line operations:

\begin{enumerate}
    \item {\tt queryDB.py -\,-obj `MWC 297' -\,-rad 10s -\,-closest True \\ -\,-getSpect True}
    \item {\tt inspectSED.py -\,-savePlt True -\,-phot MWC297/MWC297\_phot.dat \\ \tt -\,-spec MWC297/70800234\_sws.fit } \\
\end{enumerate}

The first of these operations dictates that a cone search radius of $10$\,arc seconds is to be used to identify matches in the online catalogs, flux calibrated infrared spectra will be searched for ({\tt --getSpect True}) and the closest match found in each catalog will be saved ({\tt --closest True}). Using {\tt --closest False} would instead result in an interactive session. 

When run, {\tt queryDB.py} creates a new folder in the current working directory. The name of this new folder matches the object name parsed to the {\tt --obj} argument, following removal of any spaces in the string. For MWC 297, a total of $102$ photometric data points were retrieved and saved to the newly created file, {\tt MWC297\_phot.dat}, in the newly created {\tt MWC297/} folder. With the optional argument {\tt --getSpect True} in use, a search is conducted for ISO/SWS and low resolution Spitzer spectra. In the case of MWC 297, ISO/SWS data is located. This is retrieved and also saved to the newly created {\tt MWC297/} folder, retaining the original name for the file.

The second command dictates that both the photometry and the spectroscopy retrieved using {\tt queryDB.py} should be plotted and, as optional argument {\tt -\,-savePlt True} is in use, that the plot should be saved to file. None of the data retrieved by SEDBYS have been flagged for removal. 

The full list of photometry is provided, with individual references, in Table~\ref{tab:mwc297_phot}. This was created using command \\
{\tt toLaTex.py -\,-phot MWC297/MWC297\_phot.dat}\\
which also created a corresponding bibliographic reference file.

\begin{table}
\begin{tabular}{clll}
\hline
Wavelength & $\lambda F_{\lambda}$   & Date & Reference\\
$\mu$m     & $10^{-13}$\,W\,m$^{-2}$ &      &   \\
\hline
0.36 & $0.425$ & -- & \citep{Hillenbrand1992zm} \\ 
0.36 & $0.089$ & -- & \citep{Lazareff2017lg} \\ 
0.44 & $0.416$ & -- & \citep{Hillenbrand1992zm} \\ 
0.44 & $0.480\pm0.040$ & 2006Jun09 & \citep{Tannirkulam2008jg} \\ 
0.44 & $0.418$ & -- & \citep{Lazareff2017lg} \\ 
0.44 & $0.489\pm0.011$ & -- & \citep{aavso} \\ 
0.44 & $0.532\pm0.044$ & 2006Aug27 & \citep{Tannirkulam2008jg} \\ 
0.48 & $1.006\pm0.045$ & -- & \citep{aavso} \\ 
0.51 & $3.301\pm0.035$ & -- & \citep{gaia} \\ 
0.54 & $2.383\pm0.202$ & -- & \citep{aavso} \\ 
0.54 & $2.491\pm0.184$ & 2006Jun09 & \citep{Tannirkulam2008jg} \\ 
0.54 & $2.561\pm0.189$ & 2006Aug27 & \citep{Tannirkulam2008jg} \\ 
0.54 & $2.561$ & -- & \citep{Lazareff2017lg} \\ 
0.55 & $2.657$ & -- & \citep{Hillenbrand1992zm} \\ 
0.62 & $8.141\pm0.817$ & -- & \citep{aavso} \\ 
0.64 & $11.591\pm0.041$ & -- & \citep{gaia} \\ 
0.65 & $10.826\pm0.897$ & 2006Aug27 & \citep{Tannirkulam2008jg} \\ 
0.65 & $9.965$ & -- & \citep{Lazareff2017lg} \\ 
0.65 & $10.926\pm0.906$ & 2006Jun09 & \citep{Tannirkulam2008jg} \\ 
0.69 & $20.725$ & -- & \citep{Hillenbrand1992zm} \\ 
0.78 & $24.881\pm0.277$ & -- & \citep{gaia} \\ 
0.79 & $21.244\pm1.370$ & 2006Jun09 & \citep{Tannirkulam2008jg} \\ 
0.79 & $25.075\pm1.617$ & 2006Aug27 & \citep{Tannirkulam2008jg} \\ 
0.79 & $23.946$ & -- & \citep{Lazareff2017lg} \\ 
0.88 & $34.297$ & -- & \citep{Hillenbrand1992zm} \\ 
1.23 & $114.860$ & 1989Apr & \citep{Berrilli1992fu} \\ 
1.25 & $151.556$ & -- & \citep{Hillenbrand1992zm} \\ 
1.25 & $135.394\pm2.369$ & -- & \citep{2mass} \\ 
1.25 & $120.385$ & -- & \citep{Lazareff2017lg} \\ 
1.25 & $144.735\pm10.664$ & 2006Jun04 & \citep{Tannirkulam2008jg} \\ 
1.60 & $311.187$ & -- & \citep{Lazareff2017lg} \\ 
1.60 & $278.626$ & -- & \citep{Hillenbrand1992zm} \\ 
1.60 & $289.083\pm18.638$ & 2006Jun04 & \citep{Tannirkulam2008jg} \\ 
1.63 & $288.601$ & 1989Apr & \citep{Berrilli1992fu} \\ 
1.65 & $326.315\pm63.115$ & -- & \citep{2mass} \\ 
2.15 & $565.429\pm124.987$ & -- & \citep{2mass} \\ 
\hline
\end{tabular}
\caption{Photometry retrieved for MWC~297 using {\tt queryDB.py}, flux-converted using {\tt inspectSED.py} and tabulated using {\tt toLaTex.py}. Observation dates and flux density measurement uncertainties are provided where available.}
\label{tab:mwc297_phot}
\end{table}

\begin{table}
\begin{tabular}{clll}
\hline
Wavelength & $\lambda F_{\lambda}$   & Date & Reference\\
$\mu$m     & $10^{-13}$\,W\,m$^{-2}$ &      &   \\
\hline
2.18 & $507.814$ & -- & \citep{Hillenbrand1992zm} \\ 
2.18 & $517.255$ & -- & \citep{Lazareff2017lg} \\ 
2.18 & $489.446\pm36.064$ & 2006Jun04 & \citep{Tannirkulam2008jg} \\ 
2.19 & $473.467$ & 1989Apr & \citep{Berrilli1992fu} \\ 
3.35 & $235.555\pm24.733$ & -- & \citep{wise} \\ 
3.54 & $778.413$ & -- & \citep{Hillenbrand1992zm} \\ 
3.79 & $636.974$ & 1989Apr & \citep{Berrilli1992fu} \\ 
4.29 & $684.072\pm60.797$ & -- & \citep{msx} \\ 
4.35 & $620.536\pm61.337$ & -- & \citep{msx} \\ 
4.60 & $199.646\pm3.310$ & -- & \citep{wise} \\ 
4.64 & $662.736$ & 1989Apr & \citep{Berrilli1992fu} \\ 
4.80 & $651.671$ & -- & \citep{Hillenbrand1992zm} \\ 
8.28 & $511.240\pm14.845$ & -- & \citep{msx} \\ 
8.38 & $462.494$ & 1989Apr & \citep{Berrilli1992fu} \\ 
9.69 & $319.010$ & 1989Apr & \citep{Berrilli1992fu} \\ 
11.60 & $946.876\pm216.282$ & -- & \citep{wise} \\ 
12.13 & $308.196\pm12.357$ & -- & \citep{msx} \\ 
12.89 & $242.680$ & 1989Apr & \citep{Berrilli1992fu} \\ 
14.65 & $214.459\pm12.483$ & -- & \citep{msx} \\ 
18.00 & $363.582\pm0.691$ & -- & \citep{akariirs} \\ 
21.34 & $160.433\pm8.429$ & -- & \citep{msx} \\ 
22.10 & $38.031\pm0.035$ & -- & \citep{wise} \\ 
450 & $0.903\pm0.451$ & -- & \citep{scuba} \\ 
450 & $0.097\pm0.013$ & -- & \citep{Sandell2011yn} \\ 
850 & $0.069\pm0.014$ & -- & \citep{scuba} \\ 
850 & $0.020\pm0.001$ & -- & \citep{Sandell2011yn} \\ 
1300 & $0.028$ & 1995Feb & \citep{Henning1998xw} \\ 
1300 & $0.003$ & -- & \citep{Hillenbrand1992zm} \\ 
1300 & $0.0040\pm0.0001$ & 2006Feb & \citep{AlonsoAlbi2009st} \\ 
1300 & $0.007$ & 2006Aug28 & \citep{Manoj2007yp} \\ 
1300 & $0.0055\pm0.0003$ & -- & \citep{Henning1994lo} \\ 
2600 & $0.00172\pm0.00006$ & 2006Feb & \citep{AlonsoAlbi2009st} \\ 
6917 & $0.0001$ & 2005Dec & \citep{AlonsoAlbi2009st} \\ 
13350 & $0.00006$ & 2005Dec & \citep{AlonsoAlbi2009st} \\ 
36000 & $0.0000074\pm0.0000001$ & 1990Feb11 & \citep{Skinner1993uu} \\ 
36000 & $0.0000073\pm0.0000001$ & 1990Feb11 & \citep{Skinner1993uu} \\ 
36000 & $0.0000121\pm0.0000001$ & 1991Feb07 & \citep{Skinner1993uu} \\ 
60000 & $0.000007$ & 1991Feb07 & \citep{Skinner1993uu} \\ 
\hline
\end{tabular}
\caption{Table~\ref{tab:mwc297_phot} cont.}
\label{}
\end{table}


\subsection{Adding further extant catalogs to SEDBYS database}\label{sec:eg_additions}
Using {\tt addLocal.py} relies on the user first making a comma-separated photometric data file which matches the format of existing entries in the SEDBYS photometric database. The new file is tested to ensure:
\begin{itemize}
    \item The target identifier is SIMBAD-compatible and is provided as it appears within the SIMBAD database (exceptions apply for individual components of binary or multiple systems - see Section~\ref{sec:localDB}). For example, the canonical T~Tauri star T~Tau must be listed in full as V*~T~Tau, as it appears in SIMBAD, even though the shorthand name T Tau is SIMBAD-compatible. 
    \item The observing date is in YYYYMmmDD (e.g. 2020Jan01) or YYYYMmm (e.g. 2020Jan) format.
\end{itemize}

\begin{table}
\begin{tabular}{cl}
\hline
\textbf{Entry number} & \textbf{Metadata}\\
\hline
1 & VizieR catalog identifier or path to file in SEDBYS database \\
2 & Bibcode \\
3 & Wavelengths (in m) for each photometric measurement \\
4 & Effective resolution of each telescope or interferometric array \\
5 & The column header for each photometric measurement \\
6 & The column header for the error on each photometric measurement  \\
7 & The unit (mag, mJy, or Jy) for each photometric measurement  \\
8 & Filter name for each photometric measurement \\
\hline
\end{tabular}
\caption{Catalog metadata stored in the python dictionaries within {\tt cat\_setup.py}}
\label{tab:meta}
\end{table}

Then, the catalog metadata (see Table~\ref{tab:meta}) is parsed to {\tt addLocal.py} to add the new entry to the SEDBYS photometric database. For example, using the following command would add the file called {\tt herschel\_phot.csv}, created from table~2 of \citep{Pascual16} to the SEDBYS photometry database:\\

{\tt python3 addLocal.py -\,-nam HERSCHEL1 -\,-fil herschel\_phot.csv \\
     -\,-ref `2016A\&A...586A...6P' -\,-wav 70e-6,100e-6,160e-6 \\
     -\,-res 5.033,7.190,11.504 -\,-fna F70,F100,F160 \\
     -\,-ena eF70,eF100,eF160 -\,-una Jy,Jy,Jy \\
     -\,-bna Herschel:PACS:F70,Herschel:PACS:F100,Herschel:PACS:F160}\\

In contrast, a user wishing to add a new VizieR catalog to the list of queried catalogs is only required to parse the catalog metadata to {\tt addVizCat.py}. Both scripts inspect the format of the catalog metadata parsed to them, ensuring:
\begin{itemize}
    \item The catalog or the file exists.
    \item The bibliographic reference is compliant with NASA ADS formatting.
    \item The column headers provided appear in the catalog or the file being added.
    \item The measurement units are one of ``mag'', ``mJy'', or ``Jy''.
    \item If the measurement unit is ``mag'', the filter keyword must match one of the entries in SEDBYS file {\tt zero\_points.dat} to ensure it can be correctly converted to a flux density. Otherwise, it simply needs to be descriptive (e.g. ALMA:F1300). If the filter is not recognised, the user is prompted to add the filter zero point and central wavelength to the zero\_points.dat file.
\end{itemize}

For example, the following command would ensure VizieR catalog I/259/tyc2 \citep{tycho2} is added to the list of online catalogs to be queried:\\

{\tt python3 addVizCat.py -\,-nam TYCHO2 -\,-cat `I/259/tyc2' \\
     -\,-ref `2000A\&A...355L..27H' -\,-wav 426e-9,532e-9
     -\,-res 0.37,0.462 \\ -\,-fna BTmag,VTmag -\,-ena e\_BTmag,e\_VTmag
     -\,-una mag,mag \\ -\,-bna HIP:BT,HIP:VT}

\section{Impact}
\label{}




In developing SEDBYS, I aimed to dramatically speed up and improve the process of collating archival spectro-photometric data to build SEDs for YSOs. The main advancements come in two parts: (i) the collation of a photometry database containing previously published archival data that is not catalogued in virtual observatory tables; (ii) an interactive plotting tool, designed to be used in conjunction with collated observation date and resolution information, which remains synced to the collated data files. In addition, by including scripts which allow individual users to expand the database with new and existing catalogs and/or data ensures that SEDBYS remains useful to the community for years to come. 

Three major benefits of using SEDBYS rather than other virtual observatory tools are illustrated in Section~\ref{sec:example}. Firstly, in Fig.~\ref{fig:mwc297}, the SED collated for YSO MWC 297 is shown. None of the data obtained at wavelengths longer than $850\,\mu$m is possible to retrieve using existing SED compilation tools which interface with virtual observatory tables (e.g. VizieR photometry viewer and VOSA). Instead, all of this data is retrieved from the SEDBYS photometric database. This is by no means restricted to this one YSO and is instead common for long wavelength data as flux measurements are rarely tabulated, instead being reported in-text within papers focused on either individual or relatively few objects at a time. Without SEDBYS, someone compiling an SED would have to locate each of the publications these data were retrieved from and collate the data themselves. This is particularly time consuming for students or newcomers to long wavelength observations and, even for experts, may take a few days.

Secondly, in Table~\ref{tab:mwc297_phot}, the collated photometry is shown, alongside the observation date information, where available. This observation date information is not collated by all SED compilation tools relying on virtual observatory tables (VizieR photometry viewer, for example). With SEDBYS, the user can easily trim their SED for MWC~297 to only include the $BVR_{\rm{c}}I_{\rm{c}}JHK$ photometry obtained in 2006Jun, for instance. In this way, inferences regarding the nature of the star and its circumstellar environment could be made without having to consider the effects of photometric variability. 

It is also worth noting that the \citep{Tannirkulam2008jg} study from which the contemporaneous $BVR_{\rm{c}}I_{\rm{c}}JHK$ photometry for MWC~297 was retrieved is also not available in virtual observatory table format so these data would not be retrieved if the user used VizieR or VOSA, for example. Without SEDBYS, a user wishing to inspect their collated photometry for photometric variability effects would have to first search through the literature for extant data published with observation date information before then looking up the applicable zero points for the (sometimes non-standard) photometric filters used. In this way, SEDBYS massively saves the user time, as before, but also ensures a greater legacy value of the original published data.

Thirdly, SEDBYS provides flux-calibrated low resolution infrared spectra in addition to the collated photometry. This data would otherwise have to be sourced independently, again adding to the length of time it would take the user to collate all the information they required for their analysis. 

While the SEDBYS photometry database has been being compiled, the suite of command line routines has been used in \citep{Davies18}, \citep{Labdon19} and \citep{Davies20} to compile SEDs for YSOs HD~142666, SU Aur, and RY~Tau, respectively. In particular, the extant photometry and flux measurements compiled using SEDBYS were used to obtain an independent assessment of the stellar flux contribution to the flux across near-infrared wavelengths. SEDBYS has since been adopted by the Gemini-LIGHTS (Gemini Large Imaging with Gpi Herbig/T-tauri Survey \citep{Laws20}) young stellar object survey team (Laws et al. \textit{in prep}; Davies et al. \textit{in prep}; Rich et al. \textit{in prep}). SEDBYS usefulness as a teaching/learning tool for undergraduate astronomy students has also been recognised (T. J. Harries, 2020, private communication; S. G. Gregory 2020, private communication). 

Finally, the inclusion of functions which enable the further expansion of the SEDBYS photometry database and the list of online surveys and catalogs queried by {\tt queryDB.py} makes SEDBYS an easily expandable tool. Its design has been focused on young stars but flexibility in its design allows for expansion by users interested in other astronomical research areas where stellar SEDs are a useful tool.

\section{Conflict of Interest}


No conflict of interest exists:
We wish to confirm that there are no known conflicts of interest associated with this publication and there has been no significant financial support for this work that could have influenced its outcome.

\section*{Acknowledgements}
CLD acknowledges support from ERC Starting Grant ``ImagePlanetFormDiscs'' (Grant Agreement No. 639889) and thanks Tim J. Harries, John D. Monnier, and \'{A}lvaro Ribas for helpful discussions, and Anna Laws and Evan Rich for their assistance in testing the SEDBYS tool. 
This research has made use of the SIMBAD database, operated at CDS, Strasbourg, France, and the VizieR catalogue access tool, CDS, Strasbourg, France (DOI : 10.26093/cds/vizier). The original description of the VizieR service was published in 2000, A\&AS 143, 23. 
This research has made use of NASA's Astrophysics Data System Bibliographic Services.



\bibliographystyle{elsarticle-num} 
\bibliography{sedbys.bib}






%


\end{document}